\documentclass[a4paper]{jpconf}
\usepackage{graphicx}
\begin{document}
\title{Inclusive breakup of Borromean nuclei}

\author{Mahir S. Hussein}

\address{Instituto Tecnol\'{o}gico de Aeron\'{a}utica, DCTA,12.228-900 S\~{a}o Jos\'{e} dos Campos, SP, Brazil}
\address{Instituto de Estudos Avan\c{c}ados, Universidade de S\~{a}o Paulo C. P.
72012, 05508-970 S\~{a}o Paulo-SP, Brazil}
\address{Instituto de F\'{\i}sica,
Universidade de S\~{a}o Paulo, C. P. 66318, 05314-970 S\~{a}o Paulo,-SP,
Brazil}
\author{Brett V. Carlson, Tobias Frederico}
\address{Instituto Tecnol\'{o}gico de Aeron\'{a}utica, DCTA,12.228-900 S\~{a}o Jos\'{e} dos Campos, SP, Brazil}

\ead{hussein@if.usp.br;brett@ita.br;tobias@ita.br}

\begin{abstract}
We derive the inclusive breakup cross section of a three-fragment projectile nuclei, $a = b +x_1 + x_2$, in the spectator model. 
The resulting four-body cross section for observing $b$, is composed of the elastic breakup cross section which contains information 
about the correlation between the two participant fragments, and the inclusive non-elastic breakup cross section. 
This latter cross section is found to be a non-trivial four-body generalization of the Austern formula \cite{Austern1987}, 
which is proportional to a matrix element of the form, 
$\langle\hat{\rho}_{{x_1},{x_2}}\left|\left[W_{{x_1}} + W_{{x_2}} + W_{3B}\right]\right|\hat{\rho}_{{x_1}, {x_2}}\rangle$. 
The new feature here is the three-body absorption, represented by the imaginary potential, $W_{3B}$. 
We analyze this type of absorption and supply ideas of how to calculate its contribution.
\end{abstract}

\section{Introduction}
The theory of inclusive breakup was developed in the 80's \cite{Austern1987, IAV1985, UT1981, HM1985} for a reaction induced by a 
two-fragment projectile, $a = b + x$ with $b$ taken as the detected spectator fragment, and $x$ being the interacting participant fragment.
The HM theory was applied to the alpha spectra (b = $\alpha$) using the eikonal approximation for the distorted wave in \cite{HM1989}. 
The IAV and the HM theory were based on the post form where the interaction is $V_{xb}$, while the UT theory is based on the prior form with 
the interaction given by $U_{b} + U_{x}  - U_{a}$. The HM expression for the inclusive non-elastic breakup cross section was
further greatly analyzed in \cite{Canto1998}.
In a recent 
publication \cite{Potel2015}, Potel et al. tested the Surrogate Method \cite{Escher2012} in the case of $(d, p)$ 
reaction on the actinide nuclei to be used in projected breeder reactors. For this purpose, they 
employed the theory of inclusive non-elastic breakup reactions, where the proton is treated as a 
spectator, merely scattering off the target, and the neutron is captured by the target. Other papers on the $(d, p)$
reaction were also published in 2015 dealing with the same issue \cite{Moro1-2015, Moro2-2015, Carlson2015, Ducasse2015}. 
Ref. \cite{Moro1-2015} also discussed the application of this  hybrid picture (direct breakup 
followed by compound nucleus formation of the subsystem)  to the reaction $^{6}$Li +$^{209}$Bi 
$\rightarrow \alpha + X$, at $E_{Lab}$ = 24 MeV and 32 MeV. \\

At near-barrier energies the three-body Austern formula alluded to above \cite{Austern1987} can in principle calculate 
the incomplete fusion part of the total fusion cross section. 
The extension of the Austern formula to the case of three fragment and Borromean projectiles,
such as the weakly bound stable nucleus $^9$Be = $\alpha + \alpha$ + n, 
and Borromean, two-neutron halo, nuclei like $^6$He = $\alpha$ + n + n, $^{11}$Li = $^9$Li + n + n, 
$^{14}$Be = $^{12}$Be + n + n, and $^{22}$C = $^{20}$C + n + n, is certainly important as more data 
on these reactions have become available. Data on complete fusion and total fusion around the 
Coulomb barrier are currently being obtained and analyzed using the effective two-body "four-body" 
Continuum Discretized Coupled Channels model, which is basically unable to calculate the incomplete 
fusion part of the total fusion cross section \cite{DTT2002}, requiring urgent derivation and 
developments of the the four-body inclusive breakup cross section. This is the purpose of this 
Contribution.

\section{The inclusive breakup of  two-fragments projectile cross section within the  spectator model}
The formula for the inclusive breakup of a two-fragment projectile has been derived by \cite{Austern1987} using a three-body model. The form of the cross section is,
\begin{equation}
\frac{d^{2}\sigma_b}{dE_{b}d\Omega_{b}} = \frac{d^{2}\sigma^{(EB)}_b}{dE_{b}d\Omega_{b}} + 
\frac{d^{2}\sigma^{(INEB)}_b}{dE_{b}d\Omega_{b}}
\end{equation}
The above formula was derived using a sum rule obtained by \cite{Ichimura1982,Hussein1984}. The elastic breakup is given by,
\begin{equation}
\frac{d^{2}\sigma^{(EB)}_b}{dE_{b}d\Omega_{b}} = \frac{2\pi}{\hbar v_a}\rho_{b}
(E_b)\sum_{\textbf{k}_x} \left|\langle \chi^{(-)}_{x}\chi^{(-)}_{b}|V_{bx}|\Psi_{0}^{(+)}\rangle 
\right|^2 \delta(E - E_b - E_{{\textbf{k}_x}})
\end{equation}

The inclusive non-elastic 
breakup part of the cross section has come to be known as the Austern formula which involves the 
reaction cross section of the "captured" fragment, calculated with a full three-body scattering wave 
function \cite{Austern1987}. 
\begin{equation}
\frac{\sigma^{2}_{b, INBU}}{dE_{b}d\Omega_{b}} = - \frac{2}{\hbar v_{a}}\rho_{b}(E_{b})\left\langle\hat{\rho}_{x}\left|W_{x}\right|\hat{\rho}_{x}\right\rangle \label{Austern-3B}
\end{equation}
where $\rho_{b}(E_{b}) = \mu_{b}k_{b}/ [(2\pi)^{3}\hbar^2]$, is the density of states of the observed fragment, $b$, $W_{x}$ is the imaginary part of the $xA$ optical potential, and 
$\hat{\rho}_{x}(\textbf{r}_{x}) = (\chi^{(-)}_{b}|\Psi^{(+)}_{3B}\rangle$, is the source function of the participant fragment, $x$.
At much higher deuteron  energies or other breaking up projectiles, 
researchers relied on the very simple but physically transparent Serber model \cite{Serber1950}, 
which is a natural limiting approximation of the Austern formula. 

\section{Extension to three-fragments projectiles}

As in the derivation of the inclusive breakup cross section for two-fragment projectile, we use the spectator model to calculate the inclusive breakup cross section of projectiles of the type $a = b + x_1 + x_2$, with $b$ representing the spectator, observed, fragment, while $x_1$ and $x_2$ correspond to the two participating, interacting, fragments. The result of the calculation, given in great details in \cite{CFH2016}, is,
\begin{equation}
\frac{d^{2}\sigma_b}{dE_{b}d\Omega_{b}} = \frac{d^{2}\sigma^{EB}_b}{dE_{b}d\Omega_{b}} + \frac{d^{2}\sigma^{INBU}_b}{dE_{b}d\Omega_{b}} 
\end{equation}
where the four-body elastic breakup cross section is,
\begin{equation}
\frac{d^{2}\sigma^{EB}_b}{dE_{b}d\Omega_{b}} =  \frac{2\pi}{\hbar v_{a}}\rho_{b}(E_b)\,\sum_{{\textbf{k}_{X}}} \left|\langle \chi^{3B(-)}_{X}\chi^{(-)}_{b}|[V_{{bx_{1}}} + V_{{bx_{2}}}]|\Psi_{0}^{4B(+)}\rangle \right|^2 \delta(E - E_{b} - E_{{\textbf{k}_{X}}}) \label{4BEB}
\end{equation}
and
\begin{equation}
\frac{d^{2}\sigma^{INEB}_b}{dE_{b}d\Omega_{b}} = \frac{2}{\hbar v_a}\rho_{b}(E_b) \langle \hat{\rho}_{X}|(W_{x_1} + W_{x_2} + W_{3B})|\hat{\rho}_{X}\rangle \label{CFH}
\end{equation}
with the source function $\hat{\rho}_{X}\equiv \hat{\rho}_{{x_1}, {x_2}}$ given by,
\begin{equation}
\hat{\rho}_{X}(\textbf{r}_{x_{1}},\textbf{r}_{x_{1}}) = (\chi_{b}^{(-)}|\Psi_{0}^{4B(+)}\rangle = \int d\textbf{r}_{b}(\chi_{b}^{(-)}(\textbf{r}_{b}))^{\dagger}\Psi_{0}^{4B(+)}(\textbf{r}_{b}, \textbf{r}_{x_{1}}, \textbf{r}_{x_{1}})\label{rho-4B}
\end{equation}
 Eq. (\ref{CFH}) is referred to here as the Carlson-Frederico-Hussein (CFH) cross section. The elastic breakup component of the cross section, Eq. (\ref{4BEB}), contains in the final state the distorted wave of the observed fragment, $b$, and the full scattering wave function of the $x_1 + x_2$ system in the potential $U_{{x_1}} + U_{{x_2}} + V_{{x_1}, {x_2}}$. Thus the  four-body EBU cross section contains invaluable information about the $X\equiv x_1 + x_2$ correlation both in the full incident 4B scattering wave function and the final $x_1 + x_2$ wave function, 
$\chi^{3B(-)}_{X}$. The Inclusive non-elastic breakup piece (INBU), the CFH cross section, $\frac{d^{2}\sigma^{INBU}}{dE_{b}d\Omega_{b}}$, Eq.(\ref{CFH}),   contains the four-body source function $\hat{\rho}_{X}$, the density of states of the observed $b$ fragment, and the imaginary potential of the 3B,  $x_1 + x_2 + A$, system, $W_{{x_1}} + W_{{x_2}} + W_{3B}$. with the genuine 3B imaginary potential representing the absorption of the two interacting fragments. The CFH cross section can be expressed in terms of a 4B reaction cross section as,

\begin{equation}
\frac{d^{2}\sigma^{INEB}_b}{dE_{b}d\Omega_{b}} = \rho_{b}(E_b) \sigma_{R, 4B}
\end{equation}
\begin{equation}
\sigma_{R, 4B} = \frac{2}{\hbar v_{a}} \langle \hat{\rho}_{X}|(W_{x_1} + W_{x_2} + W_{3B})|\hat{\rho}_{X}\rangle = \sigma_{R, {x_1}} + \sigma_{R, {x_2}} + \sigma_{R, 3B}
\end{equation}
where $\sigma_{R, 3B} = (2/\hbar v_{a})\langle \hat{\rho}_{X}|W_{3B}|\hat{\rho}_{X}\rangle$. This 3B absorptive potential, $W_{3B}$, is an hitherto not well studied one \cite{Hussein2004} and  corresponds to the principal result of this Contribution. The detailed derivation of this potential, as well as the interpretation of its structure are given in  \cite{CFH2016}. Its leading contribution corresponds to the excitation of the target by one of the fragments and its de-excitation by the other fragment as represented schematically in Fig. 1.

\begin{figure}[h]
\begin{center}
\includegraphics[width=25pc]{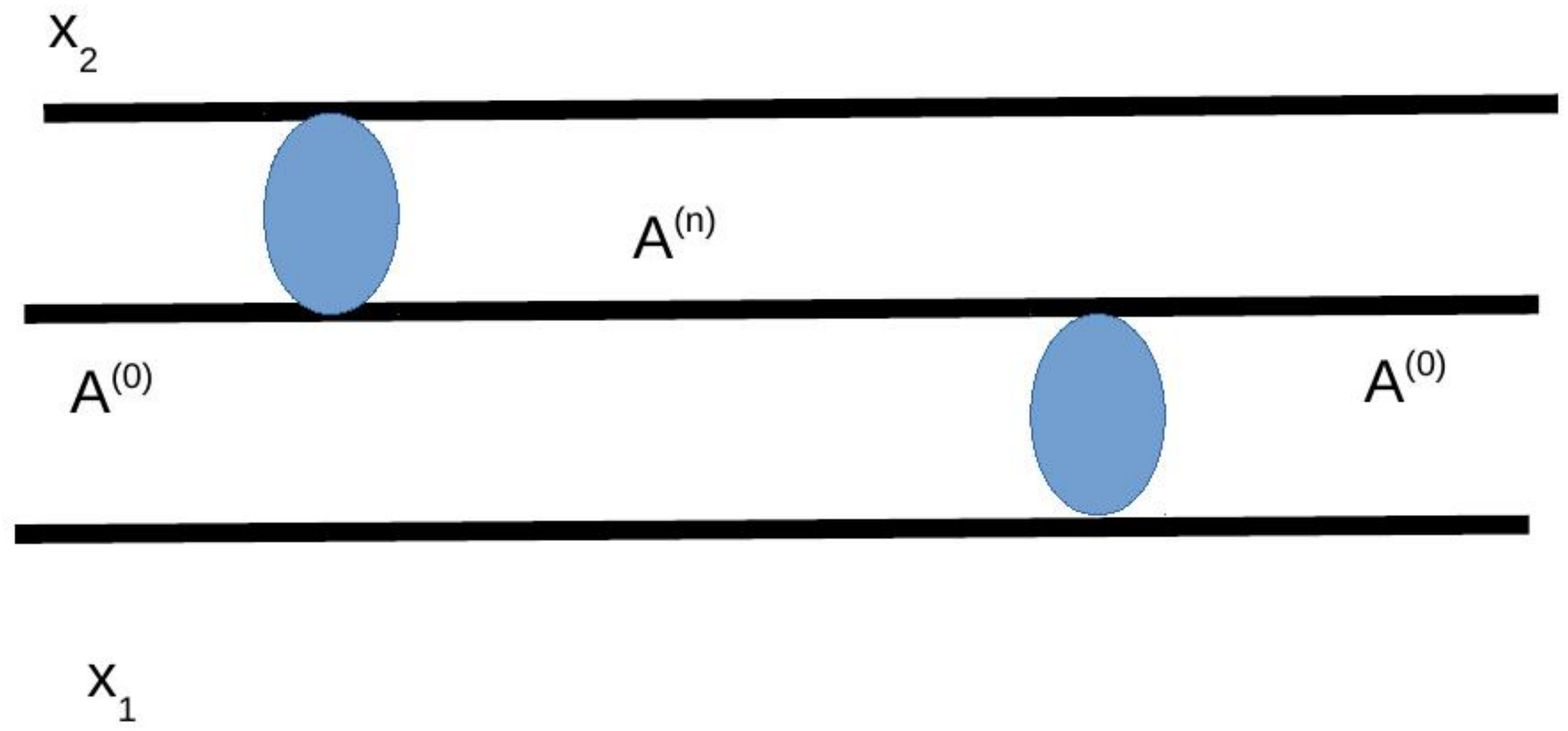}
\vspace{-1.5cm}
\caption{\label{label} Representation of the leading term in the 3-body absorptive component $W_{3B}$.}
\end{center}
\end{figure}

\section{DWBA analysis of the 4-body inclusive breakup cross section}

In the two-fragments projectiles case, the 3-body wave function was expanded in terms of distorted waves and the post interaction. The leading term came to be known as the Ichimura-Austern-Vincent (IAV) cross section \cite{IAV1985}. Using the prior form for the interaction, the leading DWBA breakup cross section was derived by Udagawa and Tamura (UT) \cite{UT1981}. Finally using the post form for the interaction but using the DWBA wave function to replace the 3-body wave function $\Psi^{(+)}_{3B}$ in Eq. (\ref{Austern-3B}), Hussein and McVoy (HM) derived their HM cross section \cite{HM1985}. It was later shown through use of the Faddeev equations and other techniques \cite{HFM1990,Ichimura1990}, that the source function $\hat{\rho}^{(+)}_{x}\approx \hat{\rho}^{(+)}_{x, IAV} = G_{x}^{(+)}(\chi^{(-)}_{b}|V_{xb}|\chi^{(+)}_{a}\phi_{a}\rangle$,
 can be written as, $\hat{\rho}^{(+)}_{x, IAV}(\bf {r}_x) = \hat{\rho}^{(+)}_{x, UT}(\bf {r}_x) + \hat{\rho}^{(+)}_{x, HM}(\bf {r}_x)$, where $\hat{\rho}^{(+)}_{x, UT}(\bf {r}_x) =  G_{x}^{(+)}(\chi^{(-)}_{b}|[U_{b} + U_{x} - U_{a}]|\chi^{(+)}_{x}\phi_{a}\rangle$, and  $\hat{\rho}^{(+)}_{x, HM} = (\chi_{b}^{(-)}|\chi^{(+)}_{a}\phi_{a}\rangle$. Accordingly the INBU cross section acquires a form composed of three terms; the UT cross section plus the HM cross section plus the interference term. The relative importance of these components was discussed in \cite{Moro2-2015}.\\

A similar DWBA analysis of the four-body cross section, the CFH equation, Eq. (\ref{CFH}),  would require the use of the Faddeev-Yakubosky equations \cite{YakSJP67} for the 4B wave function $\Psi_{0}^{4B(+)}$. The leading term in the dominant component of the FY equations would result in a the four-body equivalent of the IAV, the UT and the HM cross sections. This analysis is in progress and will be reported in a future publication. The four-body HM cross section is, on the other hand, easy to derive. It hinges on approximating the 4-body scattering wave function, $\Psi_{0}^{4B (+)}$ by the DWBA distorted wave times the ground state wave function of the projectile. The 4B inclusive non-elastic breakup theory discussed above, Eqs. (\ref{CFH}, \ref{rho-4B}), namely the CFH cross section, is potentially important for reactions involving Borromean nuclei as well in reactions involving the transfer of two neutrons and the population of Giant Pairing Vibration states \cite{Cappuzzello2015}. The capture or transfer of the two neutrons is accounted for by the 3B absorptive potential, $W_{3B}$, and the strength of this transition is measured by the value of $\sigma_{R, 3B}$. \\

Acknowledgements. We thank the Brazilian Agencies, FAPESP, CNPq and CAPES for the partial support. MSH acknowledges a Senior Visiting Professorship at ITA under the auspices of the CAPES/ITA PVS Program.


\medskip

 \begin{thebibliography}{9}
 
\bibitem{Austern1987}Austern,  Iseri N Y, Kamimura M,  Kawai M, Rawitscher G
and Yahiro M, Phys. Rep. \textbf{154}, (1987) 125  

\bibitem{IAV1985}Ichimura M, Austern N and Vincent C M, Phys. Rev. C \textbf{32}, (1985) 431

\bibitem{UT1981}Udagawa T and Tamura T, Phys. Rev. C \textbf{24}, (1981) 1348

\bibitem{HM1985}Hussein M S and McVoy K W, Nucl. Phys. A \textbf{445},  (1985) 124

\bibitem{HM1989} Hussein M S and Mastroleo R C, Nucl. Phys. A \textbf{491}, (1989) 468

\bibitem {Canto1998} Canto L F, Donangelo R, Mattos L, Hussein M S and Lotti P, Phys. Rev. C \textbf{58}, (1998) 1107


\bibitem{Potel2015}Potel G, Nunes, F M and Thompson  I  J, Phys. Rev. C \textbf{92}, (2015) 034611

\bibitem{Escher2012}Escher J E, Burke J T, Dietrich F S, Scielzo N D Thompson I J and Younes W, Rev. Mod. Phys. \textbf{84}, (2012) 353

\bibitem{Moro1-2015}Lei  J, and Moro A M, Phys. Rev. C \textbf{92}, (2015) 044616
\bibitem{Moro2-2015}Lei J and Moro A M, C \textbf{92}, (2015) 061602(R)

\bibitem{Carlson2015} Carlson B V, Capote R and Sin M, Few-Body Syst. {\bf 57}, (2016) 307

\bibitem{Ducasse2015} Ducasse Q. et al., arXiv:1512.06334 [nucl-ex]

\bibitem{Ichimura1982} Kasano A and  Ichimura M, Phys. Lett. B {\bf 115}, 1982) 81

\bibitem{Hussein1984} Hussein M S, Phys. Rev. C {\bf 30}, (1984) 1962.

\bibitem{Serber1950} Serber R, Phys. Rev. \textbf{80}, (1950) 1098

\bibitem{DTT2002} Diaz-Torres A and Thompson I J, Phys. Rev. C, \textbf{65}, (2002) 024606

\bibitem{CFH2016} Carlson B V, Frederico T and Hussein M S, arXiv:1607.05783

\bibitem{Hussein2004}Hussein M S, Carlson B V, Frederico T and Tarutina T,  Nucl. Phys. A, \textbf{738}, (2004) 367

\bibitem{HFM1990}Hussein M S, Frederico T and Mastroleo R C, Nucl. Phys. A \textbf{511}, (1990) 269

\bibitem{Ichimura1990}Ichimura M, Phys. Rev. C \textbf{41}, (1990) 834

\bibitem{YakSJP67} Yakubovsky O A, Yad. Fiz. \textbf {5}, 1312 (1967) [Sov. J. Nucl. Phys. {\textbf 5}, 1312 (1967)].

\bibitem{Cappuzzello2015} Cappuzzello F, Carbone D, Cavallaro M, Bond` M, Agodi C, Azaiez F, Bonaccorso A, Cunsolo A, Fortunato L, Foti A, Franchoo S, Khan E, Linares R, Lubian J, Scarpaci J A and Vitturi A, Nature Communications, \textbf{6}, (2015) 6743 

   



\end{thebibliography}
\end{document}